\documentclass[twocolumn,prl,aps]{revtex4}
\usepackage{graphicx}
\begin{document}


\title{Universality of the edge tunneling exponent 
of fractional quantum Hall liquids}

\author{Xin Wan$^1$}
\author{F. Evers$^1$}
\author{E.~H. Rezayi$^2$}

\affiliation{$^1$Institut f\"ur Nanotechnologie, 
Forschungszentrum Karlsruhe, 76021 Karlsruhe, Germany}
\affiliation{$^2$Department of Physics, California State University,
Los Angeles, California 90032, USA}

\date{\today}

\begin{abstract}
Recent calculations of the edge tunneling exponents in quantum Hall states
appear to contradict their topological nature. 
We revisit this issue and find no fundamental discrepancies.
In a microscopic model of fractional quantum Hall liquids with
electron-electron interaction and confinement, we calculate the edge
Green's function via exact diagonalization. 
Our results for $\nu = 1/3$ and 2/3 
suggest that in the presence of Coulomb interaction,
the sharpness of the edge and the strength of the edge confining potential,
which can lead to edge reconstruction, are the parameters that are relevant 
to the universality of the electron tunneling I-V exponent. 

\end{abstract}

\maketitle


One of the most intriguing characteristics of incompressible quantum Hall fluids is the 
nature of 
their edge excitations.  Wen\cite{wen95} has argued that Hall fluids, which have 
no order 
parameter associated with a broken symmetry, are nonetheless ordered topologically. 
While direct experimental probing of the topological order is difficult, an indirect
probe is provided by tunneling into the edge of the Hall fluid. In
fact, by virtue of the topological order, edge
 modes are uniquely determined by the physics of the bulk and,
in Abelian Hall states, form chiral Luttinger liquids\cite{wen92} (CLL).  For
tunneling from a 3-d Fermi liquid into the edge,
CLL theory leads to a non-Ohmic tunneling current voltage relation 
$I \sim V^\alpha$, 
in sharp contrast to the Ohmic prediction of a Fermi-liquid dominated edge. 

For the Hall states at $\nu=n/(2np+1)$ (where $n$ is a nonzero integer and  
$p$ is an even positive integer), the edge for $n>0$ does not contain counter propagating 
modes and the exponent is $\alpha=p+1$, independent of $n$. The situation is more complicated
for $n<0$ where counter propagating modes can be back-scattered. However, in the 
presence of disorder, the exponent takes on the universal\cite{kane94} value $\alpha=p+1-2/|n|$. 
While experiments~\cite{chang96,grayson98,chang01,hilke01,chang03} have
confirmed the nontrivial power-law behavior,
they do not agree with CLL values\cite{wen92}.
In particular, one experiment~\cite{grayson98} found an approximate 
dependence of $\alpha \approx 1/\nu$. 

Earlier attempts~\cite{han97,conti98,zuelicke99,lee98,lopez99} 
to resolve the apparent discrepancy between
experiment and theory have been summarized in 
~\cite{levitov01}.
Many of these approaches have invoked additional physics beyond
the standard theory to address the shortcoming rather than 
invalidating the basic CLL picture.
One such addition arises from the
presence of a positive background charge. On purely electrostatic grounds
the electron density near the edge 
may become quite different from that of an ideal edge~\cite{chang01}.
This effect can even lead to the reconstruction of the 
edge~\cite{wan02,yang03,joglekar03,wan03} provided the background charge is
sufficiently far from the electron layer (which is usually the case in
cleaved-edge samples). As a consequence, the tunneling characteristics
could become very sensitive to the edge profile and the universal
tunneling characteristics may not necessarily be observed. 

Meanwhile, yet another line of thought~\cite{goldman,mandal01,mandal02}
which questions the role of the
range of electron-electron interaction has emerged.
Tsiper and Goldman (TG) studied the edge wave function 
using exact diagonalization
in the presence of Coulomb interaction~\cite{goldman}.  
They concluded that the tunneling exponent depends on the range of
interaction. Crucial to their study is the assumption 
that the exponent $\alpha$ may be obtained from the ratio
of the electron occupation numbers of the two outermost occupied orbitals
for the corresponding Laughlin state in the disk geometry,
i.e. $\alpha=\rho(m^L_{\rm max}-1)/\rho(m^L_{\rm max})$.
This relation, however, has been derived only in the case of ultra short 
range interactions and its validity for the more generic finite range case 
has not been established.

Using composite fermion (CF) 
theory\cite{jain89}, Mandal and Jain\cite{mandal01,mandal02} (MJ) 
have arrived at essentially the same
conclusion. These authors adopted a hard edge by cutting off angular momentum
larger than $m_{max}=3(N-1)$ for $\nu = 1/3$ 
and, as TG, ignored the background charge.  
They found that for the ultra short range potential (which produces the Laughlin state),
the asymptotic edge Green's function exponent agreed with CLL theory.  On the
other hand, for generic potentials, in particular the Coulomb potential, a
substantial reduction of the exponent from the CLL value of 3 was observed. 
MJ attributed this reduction to the residual repulsion (beyond their hard core) 
among the composite fermions generated by the long-range 
Coulomb potential.  For $\nu=1/3$, the exponent is below 2.5 and even larger reductions 
were found for $\nu=2/5$ and $3/7$.

These results not only are at odds with the predictions 
of CLL,
but cast doubt on the most crucial element of the FQH
physics itself, namely the concept of topological order. 
The unusual properties of the chiral edge liquid is understood
to be the signature of the topological structure of the bulk 
and therefore should persist as long as the bulk exhibits the FQH
effect~\cite{wen95}. Hence, one expects the same exponent irrespective 
of the range of the interactions so long as the bulk physics remains the same.


In this paper we show that there are no fundamental contradictions with 
CLL and/or the topological order of FQH states.
In the presence of long-range Coulomb interaction, our findings suggest
that the detail of the edge confinement is relevant to understanding the 
behavior of the edge tunneling exponent.
We first address the edge exponent in a system with
long-range Coulomb interactions in the absence of neutralizing background 
charge.
To this end we evaluate the edge Green's function by exact
diagonalization in a microscopic model of the FQH liquids.  
We impose an edge confining potential by restricting the single-particle 
angular momenta to be $\le m_{\rm max}$.
We find that, for $\nu = 1/3$, the tunneling
exponent remains unchanged with Coulomb interaction for soft edges
(large $m_{\rm max}$). 
This is in sharp contrast to the reduction of $\alpha$
as found previously by MJ for hard edge confinement (small $m_{\rm max}$).
We then
investigate the effect of the edge potential induced by background
charge in the presence of long-range interaction.  For $\nu = 1/3$ and 
weak confining potential, we
again observe substantial deviations from the universal value for hard edges,
which may be highly relevant to the 
experimental studies~\cite{grayson98,chang01,hilke01}.
We also find finite-size corrections to $\alpha$ for soft edges, consistent
with the edge reconstruction scenario~\cite{wan02,yang03,wan03}. 
For $\nu = 2/3$, we find behavior consistent with strongly coupled edges
for strong confining potential and with an uncoupled outer $\nu = 1$ edge
for weak confining potential. 
Finally, we further emphasize the importance of including the edge
confining potential by comparing the ground state quantum numbers for $\nu=2/5$
obtained by exact diagonalization to those of the CF and hierarchical constructions.


We consider a microscopic model of a two-dimensional electron gas (2DEG)
confined to a two-dimensional disk 
with neutralizing background charge distributed uniformly 
on a parallel disk of radius $a$, at a distance $d$ above the 2DEG. 
The choice of $a = \sqrt{2N/\nu}$ guarantees 
that the disk encloses $N$ electrons and exactly $N/\nu$ magnetic flux quanta 
for the desired filling factor $\nu$.
The bare Coulomb interaction between the background charge and the
electrons gives rise to the confining potential. We use the same
Hamiltonian as in our previous study\cite{wan03}.
We confine the electrons to the lowest Landau level (LL) and
employ symmetric gauge wavefunctions:
$\phi_m(z) = (2 \pi 2^m m!)^{-1/2} z^m e^{-|z|^2/4}$, 
where $z = x+iy$ is the complex coordinate. In this paper the distances 
are measured in units of the magnetic length $\ell_B=\sqrt{\hbar/eB}$.

We diagonalize the Hamiltonian to obtain the exact many-body ground
state $\psi$ using the Lanczos algorithm. 
We then calculate the equal-time edge Green's function, 
\begin{equation}
G_{edge} ({\bf r} - {\bf r}') = 
\frac{\langle \psi | \Psi_e^{\dagger} ({\bf r}) \Psi_e ({\bf r}') | \psi
\rangle}{\langle \psi | \psi \rangle},
\end{equation}
where $\Psi_e^{\dagger} ({\bf r})$ and $\Psi_e ({\bf r}')$ 
are field operators, which create and annihilate an electron 
at ${\bf r}$ and ${\bf r}'$, respectively, 
on the edge of the 2DEG disk
with a radius of $R$ and $|{\bf r}-{\bf r}'|=2R\sin (\theta/2)$. 
The choice of $R$ is not crucial and will be specified later. 
In the large $|{\bf r} - {\bf r}'|$ limit, 
the edge Green's function is expected to exhibit the asymptotic
behavior
\begin{equation}
\label{scaling}
|G_{edge} ({\bf r} - {\bf r}')| \sim |{\bf r} - {\bf r}'|^{-\alpha}
\propto |\sin (\theta/2)|^{-\alpha}.
\end{equation}
Because of the relativistic invariance of CLL, the equal time and equal 
distance exponents of the Green's function are equal; 
the latter is measured in tunneling experiments.

\begin{figure}
{\centering 
\includegraphics[width=1.6in]{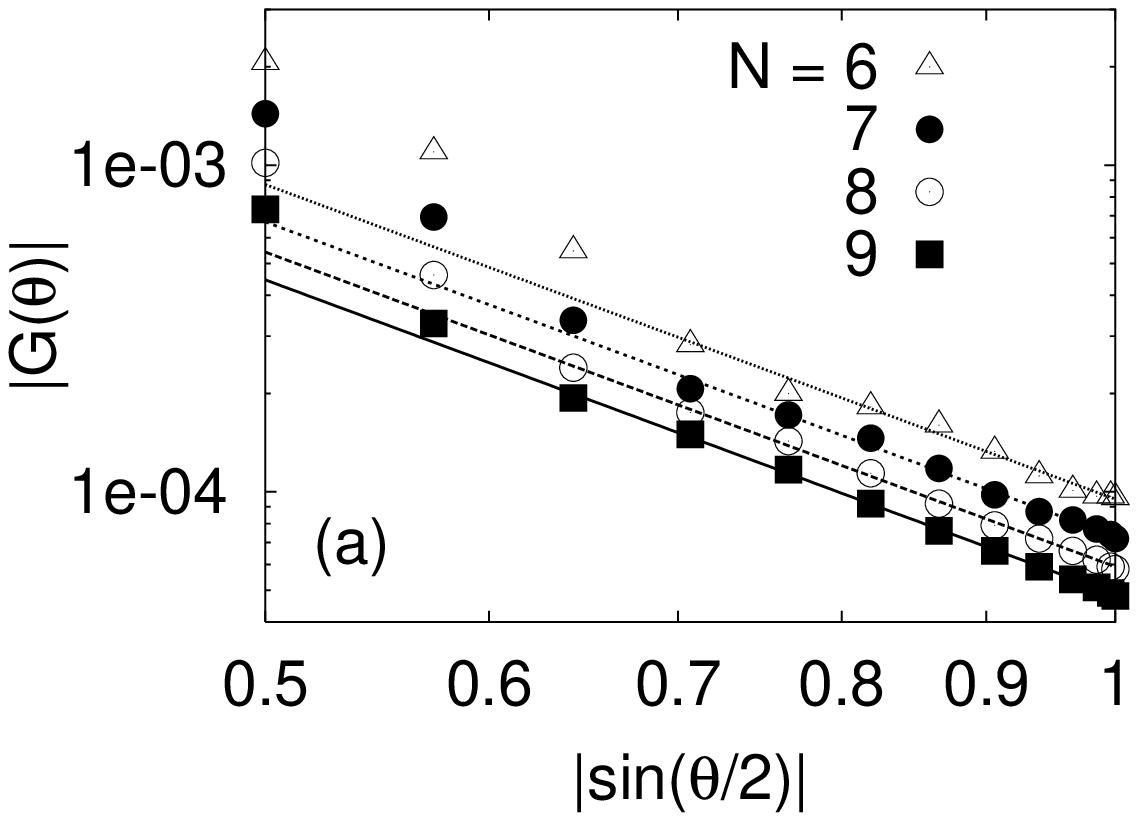} 
\hspace{-4mm}
\includegraphics[width=1.6in]{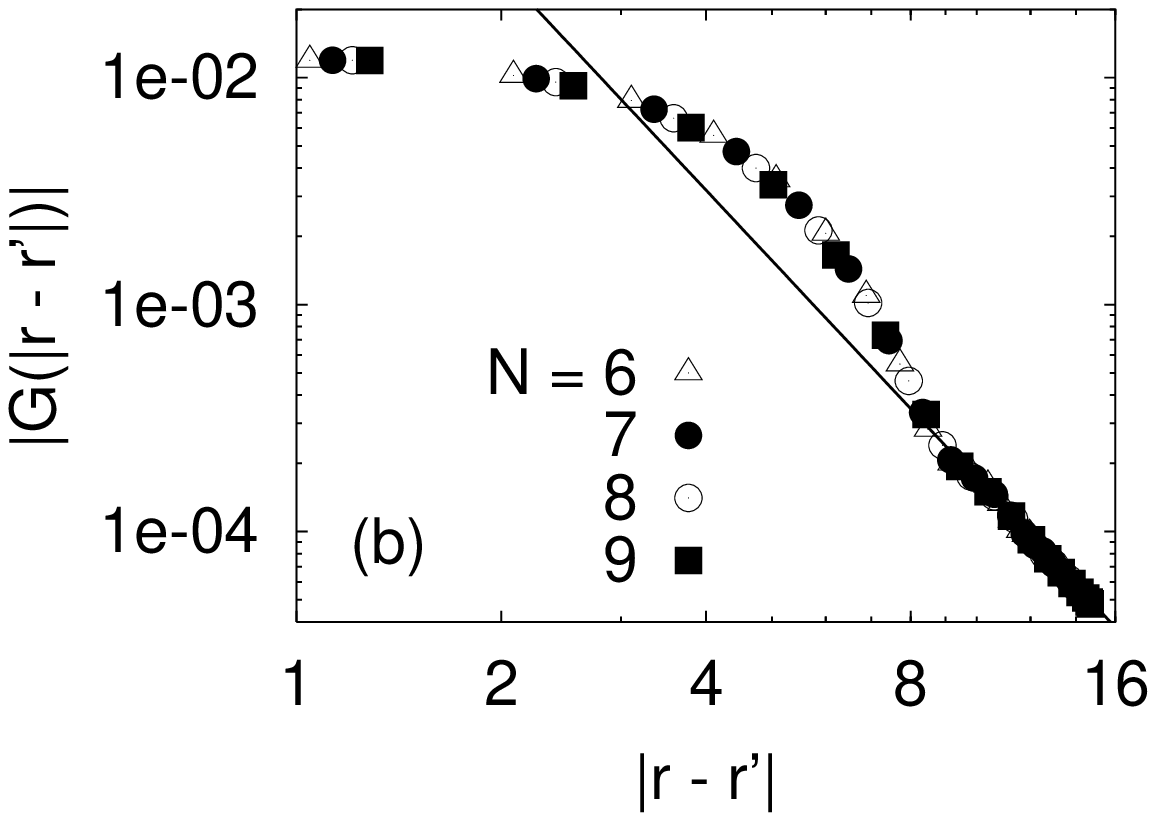} 
\includegraphics[width=1.6in]{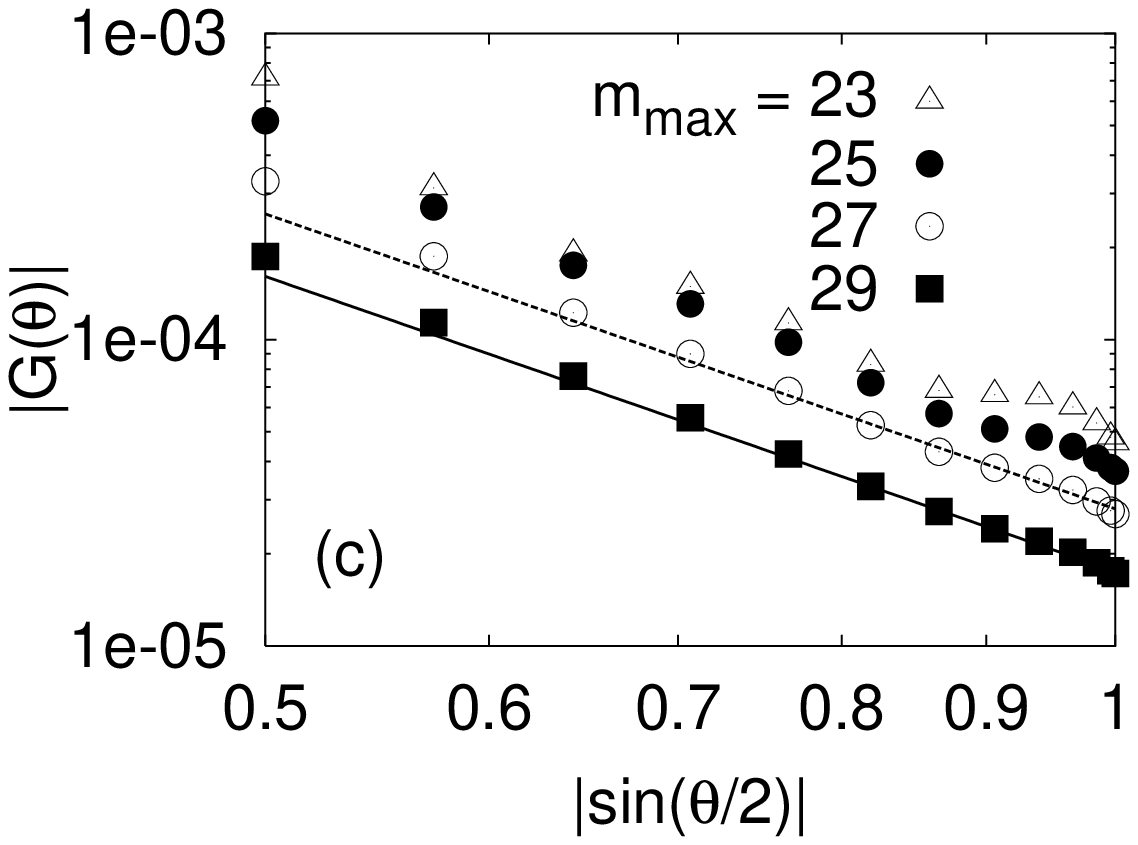} 
\hspace{-4mm}
\includegraphics[width=1.6in]{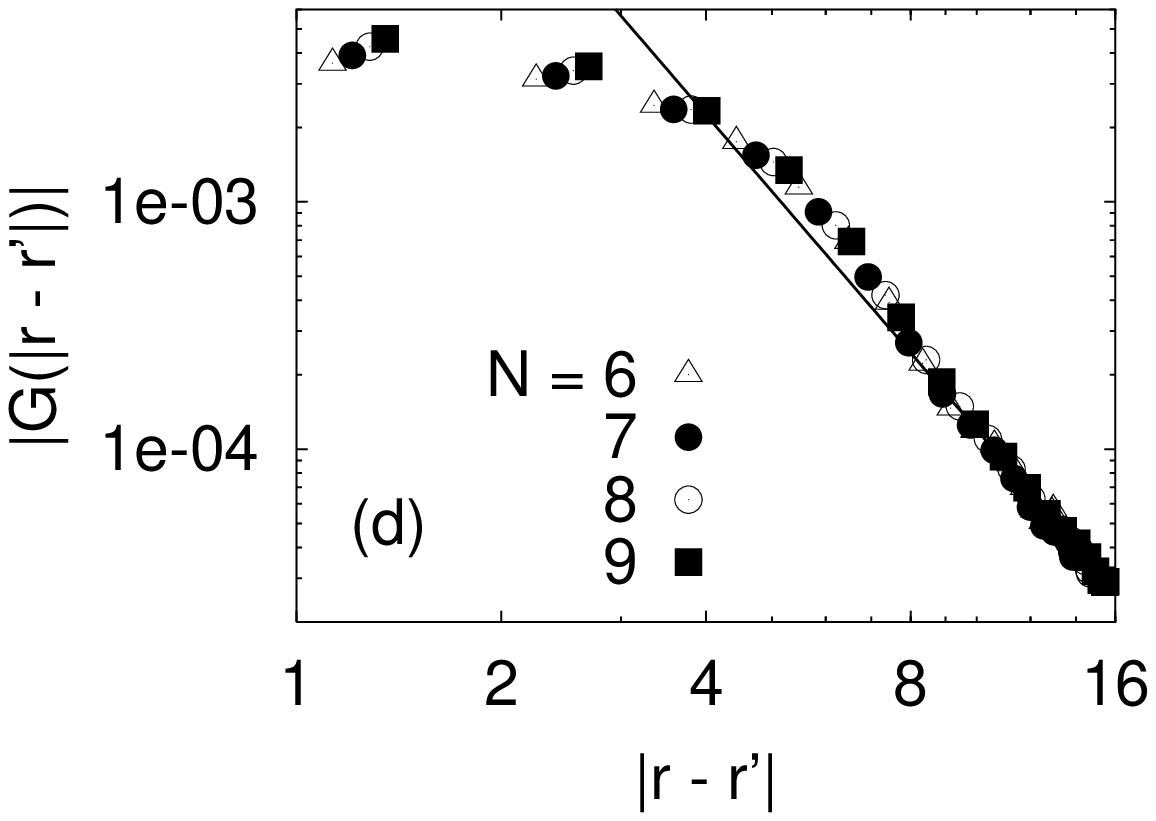} 
}
\caption{
\label{fig:compare}
The edge Green's function $|G|$ for the Laughlin state with 6-9 electrons 
at filling fraction $\nu = 1/3$
(a) as a function of $\theta$ and 
(b) as a function of $|{\bf r} - {\bf r}'|$. 
(c) $|G (\theta) |$ for 8 electrons with Coulomb interaction confined to 
orbitals with the largest angular momentum $m_{\rm max} = 23$-29. 
(d) $|G (|{\bf r} - {\bf r}'|) |$ for $N=6$-9 electrons with Coulomb
interaction confined to orbitals with $m_{\rm max} = (N-1)/\nu + 5$ 
for $\nu = 1/3$.
The lines in the log-log plots (a)-(d) correspond to a power-law
behavior with $\alpha = 3.2$.
}
\end{figure}

For comparison, we first consider the ultra short-range hardcore potential, 
for which the Laughlin state is the exact ground state. 
We do not include the background confining potential, but choose the
ground state with the appropriate total angular momentum. 
Figure~\ref{fig:compare}(a) shows the edge Green's function ($R=\sqrt{2N/\nu}$)
for the Laughlin
state with 6-9 electrons at filling fraction $\nu = 1/3$. 
We use least-square fit to 
match our data to the power-law $|G(\theta)| \sim |\sin
(\theta/2)|^{-\alpha}$ close to $|\sin
(\theta/2)| = 1$, and obtain $\alpha = 3.2\pm 0.2$. 
The errorbar reflects the dependence of $\alpha$ on system size and
range of data to fit. 
This result is in good
agreement with $\alpha = 3$ as predicted by the CLL theory. $|G(\theta)|$
for $N = 6$ shows weak oscillation around the power-law fitting curve, 
but the finite-size effects become very weak for $N \ge 7$.  
In Fig.~\ref{fig:compare}(b), we replot $|G|$ as a
function of $|{\bf r} - {\bf r}'|$. We observe perfect
scaling even for distances $|{\bf r} - {\bf r}'|$ as small as one
magnetic length, 
which is a strong indication that finite size effects are indeed negligible.

Next we consider the long-range Coulomb interaction. 
There is an important difference here with the Laughlin state so far as
the edge is concerned;  in the latter there are no occupied single-particle angular momenta 
that exceed $m_{\rm max}^L=(N-1)/\nu$.
Thus we need to enlarge our basis set and find the number 
of orbitals beyond which the properties of the system converge. 
Figure~\ref{fig:compare}(c) plots $|G_{edge}|$ for the 
Coulomb interaction and $N = 8$ electrons at filling fraction $\nu =
1/3$ for an increasing number of orbitals ($m_{\rm max} + 1$ since we label
from $m = 0$).  
We define the edge by choosing $R = \sqrt{2(m_{\rm max} + 1)}$ hereafter. 
For $m_{\rm max} < 26$ (hard edge), 
we find a weak oscillation of $|G_{edge}|$
even near the largest distance of the system. These
oscillations are probably induced by the competition between the long-range
interaction and the edge confining potential.
Similar oscillations, existing generically at other filling fractions, 
can also be observed in the electron density
profile in the presence of Coulomb interaction~\cite{goldman,wan02}.   
Therefore, fitting $|G_{edge}|$ to Eq.~(\ref{scaling}) to extract 
$\alpha$ may not produce an accurate exponent.
On the other hand,  for $m_{\rm max} > 26$ (soft edge), 
$|G_{edge}|$ can be fit very
well by the power-law with $\alpha = 3.2 \pm 0.2$, which is the same
as the ultra short-range interaction exponent.
In Fig.~\ref{fig:compare}(d), we again show a scaling plot of $|G|$ over
$|{\bf r} - {\bf r}'|$ for $N=6-9$ electrons 
and $m_{\rm max} = (N-1)/\nu + 5$ at $\nu = 1/3$, again for the Coulomb interaction. 
Even with long range interaction, the data shows good scaling with only 
small deviations at length scales below $8$.
We note that the choice of $m_{\rm max}$ here is the same as in
Ref.~\cite{goldman}. The difference in the exponent is caused by
the manner it was determined.  We have verified that the formula used 
by TG does not agree with the exponent in the Green's function.

\begin{figure}
{\centering \includegraphics[width=3.0in]{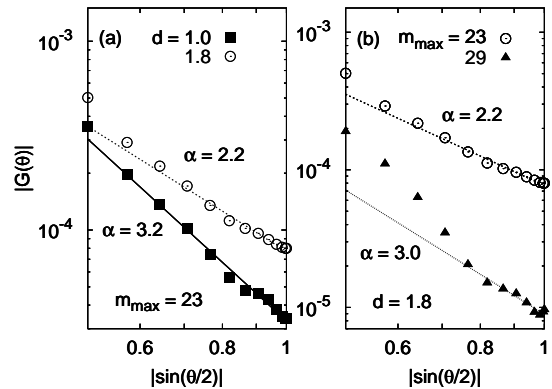} }
\caption{
\label{fig:reconstruction}
The edge Green's function $|G|$ of 8 electrons at $\nu = 1/3$ with
Coulomb interaction and the background charge confining potential for
(a) $m_{\rm max} = 23$ before ($d = 1.0$) and after the edge reconstruction
($d = 1.8$), and (b) $m_{\rm max} = 23$ and 29 with $d = 1.8$. 
The straight lines are power-law fits with $\alpha = 2.2$ and 3.2 in (a), 
and $\alpha = 2.2$ and 3.0 in (b), respectively. 
}
\end{figure}

So far we have excluded the background confining potential.
Without the background
charge, electrons tend to move to the edge to reduce their Coulomb repulsion.
This seems to induce strong density oscillations near the edge, extending into the bulk
rather than forming a roughly uniform droplet. Nor does it conform to the
experiments where a confining potential is always present.
Figure~\ref{fig:reconstruction}(a) shows the edge Green's function for
8 electrons with $m_{\rm max}= 23$ (hard edge) 
with the corresponding confining potential for $\nu = 1/3$. 
For $d = 1.0$, where
there is no edge reconstruction (strong confining potential), 
we find that 
$G(\theta)$ agrees very well with a power law of $G \sim |\sin
(\theta/2)|^{\alpha}$ with $\alpha = 3.2 \pm 0.1$. 
This is equal to the exponent in the complete
absence of any confining potential. 
This is because the background has largely mitigated the combined effects of
the long-range repulsion and the hard-edge confinement.
However, for $d>d_c\approx 1.5$, due to 
edge reconstruction, this is no longer the case.
For $d = 1.8 > d_c$ (weak confining potential), 
$G(\theta)$ increases its value as a result of electrons
moving closer to the edge and changes $\alpha$ to 
$2.2 \pm 0.1$. 
Again, one can see this qualitatively on the electrostatic
level; the electron occupation ratio 
in the lowest Landau level near the edge is larger than $1/3$. 
We next relax the cutoff in angular momentum space 
and compare $G(\theta)$ with two different $m_{\rm max}$ 
for $d = 1.8$ (Fig.~\ref{fig:reconstruction}(b)). 
For $m_{\rm max} = 29$ (soft edge), 
$g(\theta)$ shows a crossover from a power law with
$\alpha \approx 5.0$ to one with $\alpha \approx 3.0$ near $|\sin
(\theta/2)| \approx 0.75$. This suggests that the true asymptotic
behavior in the reconstructed case can only be observed at a larger
length scale. 
Such behavior agrees qualitatively with the edge reconstruction corrections
to $\alpha$ at short distances: $\delta
\alpha \propto v_\phi / v^2$~\cite{yang03,melikidze}, 
where $v_\phi$ and $v$ are velocities of
neutral and charge modes, respectively. 

The significant drop in $\alpha$ for $\nu=1/3$ in the case of hard edges 
corroborates the previous results for long-range
interaction~\cite{goldman,mandal01,mandal02}.
While it is impossible to determine with certainty what happens in the thermodynamic
limit, we agree with the assessment of MJ that these reductions are not
finite-size artifacts, notwithstanding the large distance oscillations we find in $G$. 
However, our results for soft edges appear to show that the 
non-universal behavior has more to do with the details of edge
confinement than the range of the interaction potential.
Indeed, Figs.~\ref{fig:compare} and \ref{fig:reconstruction} 
suggest that the edge confinement, through $m_{\rm max}$ as well as $d$,
is relevant to $\alpha$ in the presence of long-range interaction. 
These issues are moot for the ultra short-range interaction (unless $m_{\rm max}<(N-1)/\nu$,
in which case the Laughlin state cannot even be realized).
As pointed out in Ref.~\onlinecite{mandal01}, the CF ground state with 
one CF exciton involves only single particle states with $m\le m_{\rm max}
= 3(N-1)$, corresponding to the hard edge in our study. 
It would be interesting to find the precise CF state that would correspond to
our soft edge profile.

\begin{figure}
{\centering \includegraphics[width=3.0in]{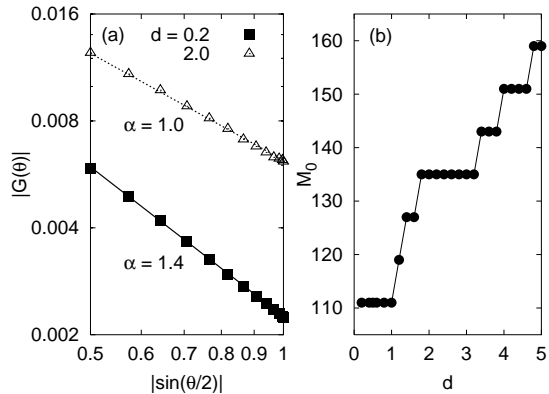} }
\caption{
\label{fig:quantumnumber}
(a) The edge Green's function $|G|$ for $\nu = 2/3$ with 
$N = 18$ and $m_{\rm max} = 27$. The straight
lines are power-law fits with exponent $\alpha = 1.4$ and 1.0 for $d = 0.2$
and 2.0, respectively. 
(b) The total angular momentum $M_0$ of the ground state as a function of
$d$, the distance between charge layers for $N = 10$ electrons at $\nu
= 2/5$ ($m_{\rm max} = 25$). 
}
\end{figure}

We have also studied the behavior of the edge Green's function at other
filling fractions, such as $\nu = 2/3$, which is not investigated in the CF
approach of MJ.
The $\nu = 2/3$ droplet can be regarded as a $\nu = 1/3$ hole droplet
superimposed on a $\nu = 1$ electron droplet. It therefore supports an
inner $\nu = 1/3$ edge and an outer $\nu = 1$ edge~\cite{macdonald90}. 
Figure~\ref{fig:quantumnumber}(a) compares the edge Green's function for
18 electrons in 27 orbitals (hard edge) with the corresponding confining
potential for $d = 0.2$ and 2.0. 
For strong confining potential ($d = 0.2$), we find, by fitting
$G(\theta)$ to a power law, that $\alpha = 1.4$  
regardless of $m_{\rm max}$. 
This is close to $1 / \nu = 1.5$ and we speculate that
the two counter propagating edge modes strongly 
couple and reconstruct into a dominant charge mode 
and a negligible neutral mode. 
On the other hand, for weak confining potential ($d = 2.0$), 
we find $\alpha = 1.0$, which probably is the fingerprint of 
the reconstructed outer edge of the $\nu = 1$ fluid. 

We have demonstrated the nontrivial effects of the edge confining
potential. We would like to emphasize that
the inclusion of the realistic confining potential not only
guarantees the charge neutrality and the homogeneity of the 2DEG, 
but also provides a numerical method of  determining 
the total angular momentum of the most stable states. 
The ground-state angular momentum $M_0$ of the interacting system 
is known for the principal filling fractions, such as $\nu = 1/3$, for
which Laughlin's variational wave function~\cite{laughlin83} 
is a good approximation. 
Based on Haldane's hierarchical construction~\cite{haldane83} 
or Jain's CF theory~\cite{jain89}, the 
variational wave function for certain 
filling fractions such as $\nu = 2/5$ can also be written down~\cite{zuelicker03,jainqdot}.
In Fig.~\ref{fig:quantumnumber}(b), we plot $M_0$   
as a function of $d$, the distance between
charge layers, for a $\nu = 2/5$ quantum Hall
liquid with $N = 10$ electrons. 
Here, $M_0$ is  determined by looking for the lowest ground state energy in
all angular momentum subspaces, since the system maintains rotational
symmetry. The nondecreasing curve is similar to those found at other filling
fractions~\cite{wan02,wan03}.
We note that $M_0$ takes a discrete value of 111, 119, 127,
and 135 for $d=0.2$-3.0, while the CF or the hierarchy construction
predicts $M_0 = 111$, 117, 125, and 135 for 3, 2, 1, and 0
CFs in the second CF Landau level (forming maximum density droplets), 
respectively~\cite{zuelicker03} (see also Ref. \cite{jainqdot}).
Evidently, the CFs in our case either form partially filled Landau levels
and/or have reconstructed edges.
We point out that the ground state with $M_0 = 135$ near $d = 2.5$ is
a $\nu = 1/3$ Laughlin state (a filled lowest Landau level of CFs).
The transition from $\nu = 2/5$ to $\nu = 1/3$
as $d$ increases is an artifact of our finite system size. Therefore,
the further increase of $M_0$ for even larger $d$ is, in fact, the
consequence of the reconstruction of the $\nu = 1/3$
edge~\cite{wan02,wan03}, albeit we set up the background charge
distribution for $\nu = 2/5$.
We note that the numerical approach applies to arbitrary $\nu$, 
which may involve many CF landau levels or be nested deep in the hierarchy. 


We thank Matt Grayson, Jainendra Jain, Akakii Melikidze, Xiao-Gang Wen and Kun Yang 
for helpful discussions. 
The work is supported by the Schwerpunktprogramme
``Quanten-Hall-Systeme'' der DFG and by US DOE under contract DE-FG03-02ER-45981.


\end{document}